\documentclass[floatfix]{revtex4}
\usepackage{amsmath}
\usepackage{color}
\usepackage{amssymb}

\usepackage{graphicx}

\begin{document}

\title{The field dependence of electron effective mass in uranium superconductors }

\author{V.P.Mineev}
\affiliation{Landau Institute for Theoretical Physics, 142432 Chernogolovka, Russia}

\begin{abstract}

The  magnetic field dependence of electron effective mass  is an  important subject
in consideration of physical properties of  superconducting ferromagnets URhGe, UCoGe and  paramagnet UTe$_2$.
It is usually accepted to find this dependence from the field dependence of $A$ coefficient in the low temperature resistivity law $\rho=\rho_0+AT^2$. 
It is shown here that the field dependence $A(H)$ takes place also
for  field-independent effective mass.
Whether or not the effective mass depends on the magnetic field can be checked by the measurements of the amplitude of de Haas - van Alphen magnetisation oscillations.

\end{abstract}

\date{\today}
\maketitle

The standard mechanism suppressing superconducting state with triplet pairing is the orbital depairing caused by magnetic field.
In addition the intensity of pairing  itself 
can be magnetic field dependent decreasing or increasing with field depending of field direction. The latter possibility  violates simple monotonic decrease of critical temperature and can lead to the peculiar phenomenon of reentrant superconductivity. Such type situation is realised 
in uranium ferromagnetic superconductors URhGe, UCoGe   for the field direction parallel to b-axis perpendicular to the spontaneous magnetisation \cite{Aoki2019}.
The former possibility is realised in UCoGe for field parallel to the spontaneous magnetisation and reveals itself as the upward curvature of the upper critical field temperature dependence \cite{Wu2017}.

In the strong coupling Eliashberg theory of superconductivity the critical temperature depends from electron effective mass renormalisation
$m^{\star}=m(1+\lambda)$  due to electron-phonon coupling according to McMillan formula $T_{sc}=\omega_D\exp\left (-\frac{1+\lambda}{\lambda}\right)$. And if the effective mass occurs magnetic field depending one can expect to get the desirable field dependent intensity of pairing interaction.  This type of field dependence of intensity of pairing has been 
proposed in \cite{Miyake2008} and then has been used in numerous publications where magnitude of $\lambda({\bf H})$ has been extracted from experimentally known upper critical field temperature dependence  \cite{Wu2017}. For  application to the ferromagnetic URhGe, UCoGe see review \cite{Aoki2019} and to paramagnetic UTe$_2$  review article \cite{Aoki2022}. However, it has been remained completely unclear {\bf why} the physics of pairing interaction in uranium superconductors is described by the   theory valid for electron-phonon coupling  but with the field dependent parameter 
$\lambda({\bf H})$.

Alternative approach to the field dependence of pairing intensity has been proposed by Hattory and Tsunetsugu \cite{Hattori2013} and 
in the series of papers by the author \cite{Mineev2011,Mineev2017,Mineev2020,Mineev2021}.
It is related not  to the field dependence of effective mass but to the field dependence of pairing interaction of electrons through the fluctuations of magnetisation mostly localised on the uranium ions. It should be noted that in frame of this approach the effective mass also depends on field  but much weaker than the field dependence of the pairing interaction itself \cite{Mineev2020}.
This mechanism has been successfully confirmed by the correct description of critical temperature behaviour for different directions of magnetic field. See, for example, the most recent publication  about a tuning of the
pairing interaction in superconducting UCoGe by magnetic field \cite{Ishida2021}. 

 It is plausible that the same mechanism determines the pairing interaction in UTe$_2$. Indeed, the recently reported  NMR results  in field applied along $b$-crystallographic axis by Y.Tokunaga et al  \cite{Tokunaga2023} demonstrate the strong increase of intensity of longitudinal magnetic fluctuations  accompanied by growth of fluctuations in perpendicular direction.
 The intensity of longitudinal fluctuations is definitely much stronger than the intensity of transverse fluctuations.
However, the relative increase of longitudinal and relative increase of transverse fluctuations under magnetic field 
are of the same magnitude. The stimulation of superconductivity in the high field region by enforce of magnetic fluctuations 
(if it is the case ) is determined by the relative increase of them in respect of zero field value. 
  This looks as a serious hint on the reason of appearance of reentrant superconducting state in strong field along  $b$-crystallographic direction. The microscopic mechanism of  fluctuations stimulation  by field   is unknown. One needs more theoretical efforts in this direction similar to the last year  ab-initio calculation of anisotropic susceptibility in UTe$_2$ successfully allowed to explain the phenomenon of the Shottky anomaly \cite{Khmelevsky2022}.

Regardless of whether affects or does not affect
the change in the effective mass with the field on the pairing intensity, there is {\bf a question},  whether the effective mass actually depends on the magnetic field {\bf ? }

Experimentally the effective mass field dependence is established by the measurements of field dependence of coefficient $A$ in low temperature behaviour of resistivity
\begin{equation}
\rho=\rho_0+\rho_{ee}=\rho_0+AT^2,
\end{equation}
where the second term originates from inelastic electron-electron scattering. The coefficient $A$ is usually proportional to square of electron effective mass
\begin{equation}
A\propto (m^{\star})^2,
\label{2}
\end{equation}
what means the fulfilment of the following relationship between the low temperature electron specific heat and $A$ coefficient
\begin{equation}
C/T\propto\sqrt{A}
\end{equation}
known as the Kadowaki-Woods relation.

So, if the field dependence of coefficient $A({\bf H})$ is experimentally established, then it would seem
it should be interpreted as the field dependence of electron effective mass $m^{\star}({\bf H})$. 
One must stress, however, that a determination of effective mass field dependence making use the Eq.(2) is indirect.
Also, concerning the field dependence of effective mass extracted from the low temperature specific heat by means  the relation $C(T)/T=\gamma({\bf H})$ one can note  that it is not always determined by the contribution of the density of states of itinerant current carriers. For instance, in the vicinity of metamagnetic transition the linear behaviour of specific heat is a general thermodynamic property.  See Eq.(33) in the paper \cite{Mineev2021}.

Still, there is a question: if the field dependence of $A$ coefficient is not due to effective mass then what is another source of its field dependence ?
Let us look on this more attentively. We begin with the derivation of the property given by Eq.(2).

The contribution to resistivity due to electron-electron scattering is 
\begin{equation}
\rho_{ee}\approx\frac{m^{\star}}{e^2n\tau_{ee}},
\end{equation}
where $n$ is the electron gas density, $e$ is the electron charge and $1/\tau_{ee} $ is the rate of electro-electron scattering. The latter is
\begin{equation}
\frac{1}{\tau_{ee}}\approx W\left (\frac{T}{\varepsilon_F}\right )^2.
\end{equation}
Here $(T/\varepsilon_F)^2$ is the probability of collision of electron pairs in the thermal layer near the Fermi surface  multiplied on the probability of this process per unit of time
$W$  which is in accordance with the Fermi-golden rule is given by the product of the square of the matrix element of electron-electron interaction $V({\bf r})$
with $\delta$ function corresponding to the energy conservation of colliding electrons integrated overall the final momenta of colliding electrons. It can be estimated as follows
\begin{equation}
W\approx \frac{ V^2}{\varepsilon_F}.
\end{equation}
 Gathering together Eqs. (4)-(6) (we put everywhere $\hbar=1$) we obtain
\begin{equation}
\rho_{ee}\approx\frac{m^{\star}}{e^2n}\frac{V^2}{\varepsilon_F}\left (\frac{T}{\varepsilon_F}\right )^2.
\end{equation}
The amplitude $V$ of  electron-electron interaction in metals  is with a good accuracy equal to the Fermi energy and we finally come to
\begin{equation}
\rho_{ee}\approx\frac {m^\star}{e^2n}\frac{T^2}{\varepsilon_F}\propto\frac {(m^\star)^2}{e^2n^{5/3}}T^2.
\end{equation}

The presented derivation is based on the assumption that electron-electron interaction completely determined by direct Coulomb  mechanism.
In magnetic materials  there is another mechanism of resistivity
 due to interaction of electrons through the magnetisation fluctuations. Then  in presence such an additional channel the scattering rate acquires the following form (Matissen rule)
\begin{equation}
\frac{1}{\tau_{ee}}+\frac{1}{\tau_M}.
\end{equation}
Below the Curie temperature the temperature dependence of electron-electron scattering rate in absence of magnetic field
is
\begin{equation}
\frac{1}{\tau_M}\propto aT^2.
\end{equation}
See, for example the paper \cite{Roy2018} where was demonstrated that the $T^2$ dependence of $\rho$ in the FM state due to the electron-electron magnetic scattering dominates over the Coulomb electron-electron scattering. This fact is in a good agreement with theoretical results \cite{Mannari1959,Ueda1975,Kaul2005} obtained in frame  $s$-$d$ model. The $a$ coefficient according to \cite{Kaul2005} is 
\begin{equation}
a\propto \frac{J_q^2 N_{0s}N_{0d}\mu_BM}{(\Delta E)^2},
\end{equation}
where $J_q$ is the Fourier component of amplitude of $s-d$ interaction,  $N_{0s}$ and $N_{0d}$ are the density of states od $s$ and $d$ electrons correspondingly, $M$ is the  magnetisation, and $\Delta E$ is the width of energy band of magnetic excitations.

The corresponding theory for orthorhombic ferromagnetic and paramagnetic compounds with $f$-electrons is not available. However, it is obvious that the scattering rate due electron-magnon interaction depends on magnetic field 
\begin{equation}
\frac{1}{\tau_M(H)}\propto a(H)T^2.
\end{equation}
through the band splitting, nonlinear dependence of magnetic susceptibility and the field dependence of the Curie temperature as it is in URhGe and UCoGe. 

Thus, the proportional to $T^2$ term becomes field dependent even in the case of field independent effective mass
\begin{equation}
\rho_{ee}\approx\frac {m^{\star}}{e^2n^{1/3}}\left (\frac{V^2}{\varepsilon_F^3}+a(H)\right )T^2.
\end{equation}

In summary, the electron-electron interaction by means of exchange of magnetic fluctuations serves as the source of the magnetic field dependence of low temperature resistivity behaviour independently from field dependence of effective mass of current carriers. Although calculation of the amplitude of contribution a(H) to resistivity through the electron-magnon interaction in the orthorhombic ferro and paramagnets with magnetic moment mostly concentrated on the uranium ions is at the moment absent, however, the problem whether the effective mass of electrons depends or does not depend on magnetic field can be solved experimentally. If the effective mass indeed is field dependent then this field dependence should reveal itself in the behaviour of amplitude of the de Haas - van Alphen magnetisation oscillations as it was demonstrated in CePd$_2$Si$_2$ and CeCoIn$_5$ [18]. For the best of my knowledge, such type direct measurements has not been performed in UCoGe and URhGe. The recent dHvA experiments on UTe2 do not reveal a effective mass field dependence\cite{Aoki-Haga2022,Eaton2023,Broyles2023}. The attraction attention to this problem is the main purpose of these remarks.

I am quite indebted to K.Ishida, Y.Tokunaga and D.Aoki for the quick and constructive reaction on my reflections on this topic.

\end{document}